\documentclass[twocolumn, apl, amsmath, amssymb, showpacs]{revtex4-2}
\usepackage{epsf}      
\usepackage{graphicx}
\usepackage{color}
\usepackage{soul}
\usepackage{gensymb}
\usepackage{sidecap}
\usepackage{amsmath}
\usepackage{mathtools}

\begin{document}

\title{Stability of the first-order character of phase transition in HoCo$_2$}
\author{Ajay Kumar}
\email{ajay1@ameslab.gov}
\affiliation{Ames National Laboratory, U.S. Department of Energy, Iowa State University, Ames, Iowa 50011, USA} 

\author{Anis Biswas}
\affiliation{Ames National Laboratory, U.S. Department of Energy, Iowa State University, Ames, Iowa 50011, USA} 

\author{Yaroslav Mudryk}
\affiliation{Ames National Laboratory, U.S. Department of Energy, Iowa State University, Ames, Iowa 50011, USA}

\date{\today}

\begin{abstract}

HoCo$_2$ exhibits a giant magnetocaloric (MC) effect at its first-order magnetostructural phase transition around 77~K, and understanding the thermodynamic nature of this transition in response to external magnetic fields is crucial for its MC applications. In this study, we present a comprehensive investigation of specific heat and magnetization measurements of HoCo$_2$ under varying magnetic fields. The specific heat measurements qualitatively indicate a transformation from first- to second-order behavior of this phase transition at higher magnetic fields.  However, analysis of the power-law dependence of the magnetic entropy change ($\Delta S_{\rm M} \propto$ H$^n$) and the breakdown of universal behavior in the temperature dependence of $\Delta S_{\rm M}$ suggest that the first-order nature remains intact, even up to 7 T. This stability of the first-order nature is further manifested through the distinctive non-linear behavior of modified Arrott plots, with a negative slope in the 6--7 T range.

\end{abstract}

\maketitle

\section{\noindent ~Introduction}

The Laves phase RCo$_2$ (R = rare earth) compounds have attracted considerable attention over the last several decades due to their intriguing phase transitions and complex magnetic properties \cite{Gratz_JPCM_01, Khmelevskyi_JMCM_2000, Stein_JMS_21}. Among all RCo$_2$ compounds, only three--HoCo$_2$, DyCo$_2$, and ErCo$_2$ exhibit first-order magnetostructural transitions, whereas the magnetic transitions of the others are of the second-order type \cite{Khmelevskyi_JMCM_2000, Inoue_JPFMP_88, Albillos_PRB_06}. However, some theoretical studies suggest that the transitions in the R = Pr and Nd compounds are also first-order \cite{Forker_PRB_03, Forker_PRB_07}, while others report a second-order nature for these  transitions \cite{Albillos_PRB_06, Hauser_PRB_98, Khmelevskyi_JMCM_2000, Parra_JalCom_09}.  The phase transitions in RCo$_2$ compounds are associated with the induction of the Co moment, resulting from the internal field of the rare earth ions, and their coupling with the moments of the R atoms \cite{Bloch_PRB_70, Gignoux_PRB_76, Berthier_JMMM_86, Moon_JAP_65, Kirchmayr_JMMM_78}. The magnetic moment of Co can vary from $\sim$1~$\mu_B$/atom for magnetic rare earth ions, where the effective internal field (H$_{\rm eff}$) acting on the Co 3$d$ electrons is higher than the critical field (H$_{\rm C}$) required to polarize the Co moments ($\sim$100 T), to zero for non-magnetic rare earth ions (Y, Lu, or Sc) \cite{Kirchmayr_JMMM_78}. Therefore, the first-order transition in R = Dy, Ho, and Er can be transformed to a second-order transition by doping with non-magnetic rare earth ions, which decreases H$_{\rm eff}$ and thus hinders the polarization of the Co moments \cite{Forker_JPCM_06, Hauser_PRB_2000}. Furthermore, the Co moment can also be induced by applying a very high external magnetic field (H $>$ H$_{\rm C}$) in the case of non-magnetic R atoms (R = Y and Lu), resulting in a metamagnetic transition \cite{Goto_SSC_89, Goto_JMMM_90, Sakakibara_PB_94, Goto_JAP_94}. The relative alignment of Co and R moments is governed by the competition between intersublattice and Hund's exchange energies. For example, the magnetic moment of Co aligns parallel to the rare earth moments for lighter atoms (R = Pr, Nd, Sm), whereas it aligns antiparallel for heavier atoms (R = Tb, Dy, Ho, Er) \cite{Moon_JAP_65, Hendy_PSS_78}.\par 

All RCo$_2$ compounds crystallize in the MgCu$_2$-type (C15) cubic structure (space group $Fd\overline{3}m$) at room temperature, which may or may not undergo a structural transition at $T_{\rm C}$, depending on the nature of the R ions \cite{Stein_JMS_21, Gratz_JPCM_01}. The R = Nd and Ho compounds also undergo an additional structural transition at the spin-reorientation temperature (T$_{\rm SR}$), which is lower than $T_{\rm C}$ \cite{Gratz_SSC_83, Xiao_JalCom_06, Mudryk_JMCC_91, Aubert_SSC_78}. Both transitions in NdCo$_2$ are second-order in nature; however, interestingly, HoCo$_2$ under ambient conditions exhibits first-order and second-order transitions at high and low temperatures, respectively \cite{Levitin_JMMM_90, Khmelevskyi_JMCM_2000}. The non-zero Co moments ($\mu_{\rm Co} \approx 1 \mu_B$) in HoCo$_2$ align antiparallel to the Ho moments ($\mu_{\rm Ho} \approx 9.5 \mu_B$) below $T_{\rm C} \sim$76~K (first-order) with an easy magnetization axis along $<$100$>$, accompanied by a structural transformation from cubic to tetragonal symmetry \cite{Gignoux_PRB_75, Moon_JAP_65, Rubinstein_JMMM_81}. Upon further cooling, HoCo$_2$ undergoes a tetragonal-to-orthorhombic transition at $T_{\rm SR} \approx 15$~K (second-order) with the easy magnetization axis reoriented along $<$110$>$ \cite{Gignoux_PRB_75, Aubert_SSC_78}. The nature and behavior of the low-temperature (LT) transition at T$_{\rm SR}$ have been extensively studied through detailed magnetization, specific heat, and temperature- and field-dependent x-ray diffraction measurements \cite{Mudryk_JMCC_91, Aubert_SSC_78}. However, an abrupt change in magnetization, electronic transport, and thermal properties has been observed at the high-temperature (HT) transition \cite{Gratz_JPCM_01, Gratz_JPCM_95, Steiner_JPFMP_78, Hauser_PRB_98, Syshchenko_JalCom_01, Albillos_PRB_06}, resulting in a giant magnetic entropy ($\Delta S_{\rm M}$) and adiabatic temperature change ($\Delta T_{\rm ad}$), which are required for magnetocaloric applications \cite{Albillos_PRB_06, Bykov_Jalcom_24, Nikitin_Cryogenics_91, Oliveira_PRB_02}. This abrupt change in various thermodynamic properties at $T_{\rm C}$ in HoCo$_2$ is associated with the first-order nature of the paramagnetic-to-ferrimagnetic phase transition. Therefore, it is essential to investigate the possible change in the nature of this transition in response to external stimuli, such as hydrostatic pressure, chemical pressure (doping), and/or magnetic field, to tailor these materials for practical use in device applications \cite{Syshchenko_JalCom_01, Hauser_PRB_98, Forker_JPCM_06, Duc_PB_89, Balli_conf_07}. \par

The change from a first-order to a second-order phase transition at T$_{\rm C}$ in HoCo$_2$ has been well established through the substitution of other rare earth ions at the Ho site \cite{Forker_JPCM_06, Balli_conf_07, Duc_PB_89}. Interestingly, Syshchenko $et$ $al.$ demonstrated that this transition can also be transformed to second-order by applying hydrostatic pressure above a critical value, P$_{\rm C}$ $\sim$3~GPa \cite{Syshchenko_JalCom_01}. This is because the ferrimagnetic (tetragonal) phase is the high-volume phase, meaning that the unit cell volume increases abruptly below T$_{\rm C}$ \cite{Mudryk_JMCC_91, Levitin_JMMM_90}. As a result, the applied hydrostatic pressure restrains this transformation, leading to a decrease in both T$_{\rm C}$ and the sharpness of the transition \cite{Syshchenko_JalCom_01, Hauser_PRB_98}. Various theoretical models have been proposed to explain this pressure-induced change in the order of the magnetic transition in HoCo$_2$ by accounting for the modification in the itinerant $d$ electrons. The applied hydrostatic pressure was found to increase the critical field required for the metamagnetic transition (H$_{\rm C}$) of the Co moments \cite{Inoue_JPFMP_88, Yamada_JMMM_95}. Moreover, P$_{\rm C}$ decreases with temperature \cite{Yamada_JMMM_95}, resulting in a complex relationship between P$_{\rm C}$, H$_{\rm C}$, and T$_{\rm C}$ in achieving the first-order transition in HoCo$_2$. On the other hand, T$_{\rm C}$ increases monotonically with an increase in the external magnetic field \cite{Mudryk_JMCC_91, Bykov_Jalcom_24}. The abrupt change in magnetization and the sharpness of the peak in the specific heat data at T$_{\rm C}$ gradually decrease with increasing magnetic field, suggesting the possibility of a field-induced change in the thermodynamic nature of this phase transition in HoCo$_2$ \cite{Mudryk_JMCC_91}. More strikingly, the discontinuity in the lattice parameters of HoCo$_2$ at T$_{\rm C}$ significantly diminishes with the application of even a 3~T magnetic field \cite{Mudryk_JMCC_91, Bykov_Jalcom_24}, supporting the notion of a field-induced change in the order of the HT transition. However, previous calorimetric measurements claim the persistence of the first-order nature of this transition in HoCo$_2$ up to 5~T \cite{Albillos_PRB_06}. Therefore, it is vital to investigate the influence of external magnetic fields on the thermodynamic nature of the HT transition in HoCo$_2$.\par

The presence of latent heat at the phase transition can be directly estimated from specific heat measurements. However, a quantitative analysis is necessary to firmly establish the thermodynamic nature of the transition. In this report, we present detailed specific heat and magnetization measurements up to 7~T to investigate the field-induced changes in the first-order nature of the transition in HoCo$_2$ at T$_{\rm C}$. Although the specific heat measurements show a suppression in the peak value and a reduction in latent heat at T$_{\rm C}$, a quantitative analysis of the magnetization data demonstrates the persistence of the first-order character of the magnetic transition in HoCo$_2$ up to at least 7~T. Based on the magnetization measurements, we employ three distinct approaches to investigate the stability of the first-order nature of this phase transition in the high-field regime.\par

\section{\noindent ~Experimental}

The polycrystalline HoCo$_2$ sample was synthesized by arc melting high-purity Ho ($>$99.7 at.\%) procured from the Materials Preparation Center of Ames National Laboratory and Co ($>$99.9\%) metal purchased from Alfa Aesar, in an argon (Ar) atmosphere. An additional 3 wt.\% Ho was used to compensate for Ho evaporation during melting. Prior to sample preparation, a Zr getter was melted to minimize the presence of residual or impurity gases in the chamber. The sample was flipped and remelted 4-5 times to ensure homogeneity. The x-ray diffraction (XRD) and energy dispersive x-ray spectroscopy (EDS) measurements (not shown here) indicate the formation of the single phase MgCu$_2$-type structure with the targeted stoichiometry. Temperature- and magnetic field-dependent magnetization measurements were conducted using a superconducting quantum interference device (SQUID) system from Quantum Design, USA (model MPMS XL-7). The temperature dependence of magnetization was measured using both temperature-stable (no overshoot) and sweep modes to elucidate the nature of the magnetic transition in this compound. The SQUID was also utilized to record field-dependent magnetization isotherms in the first quadrant at various temperatures across the magnetic transition. The magnetic field was increased from 0 to 7~T in ``no overshoot" mode to obtain high-precision data, then reduced from 7~T to 0.05~T in linear mode, and finally from 0.05~T to 0~T in oscillatory mode before moving to the next temperature (in heating mode) to avoid remanence in the sample, given its low coercivity ($\sim$10~Oe). Temperature- and field-dependent specific heat measurements were performed using a Physical Property Measurement System (PPMS) from Quantum Design, Inc., employing both the standard 2$\tau$ relaxation technique and the long heat pulse method. Further details about the measurement protocols for both magnetization and specific heat are provided in the corresponding discussions.

\section{\noindent ~Results and discussion}

\begin{figure}
\includegraphics[width=3.5in]{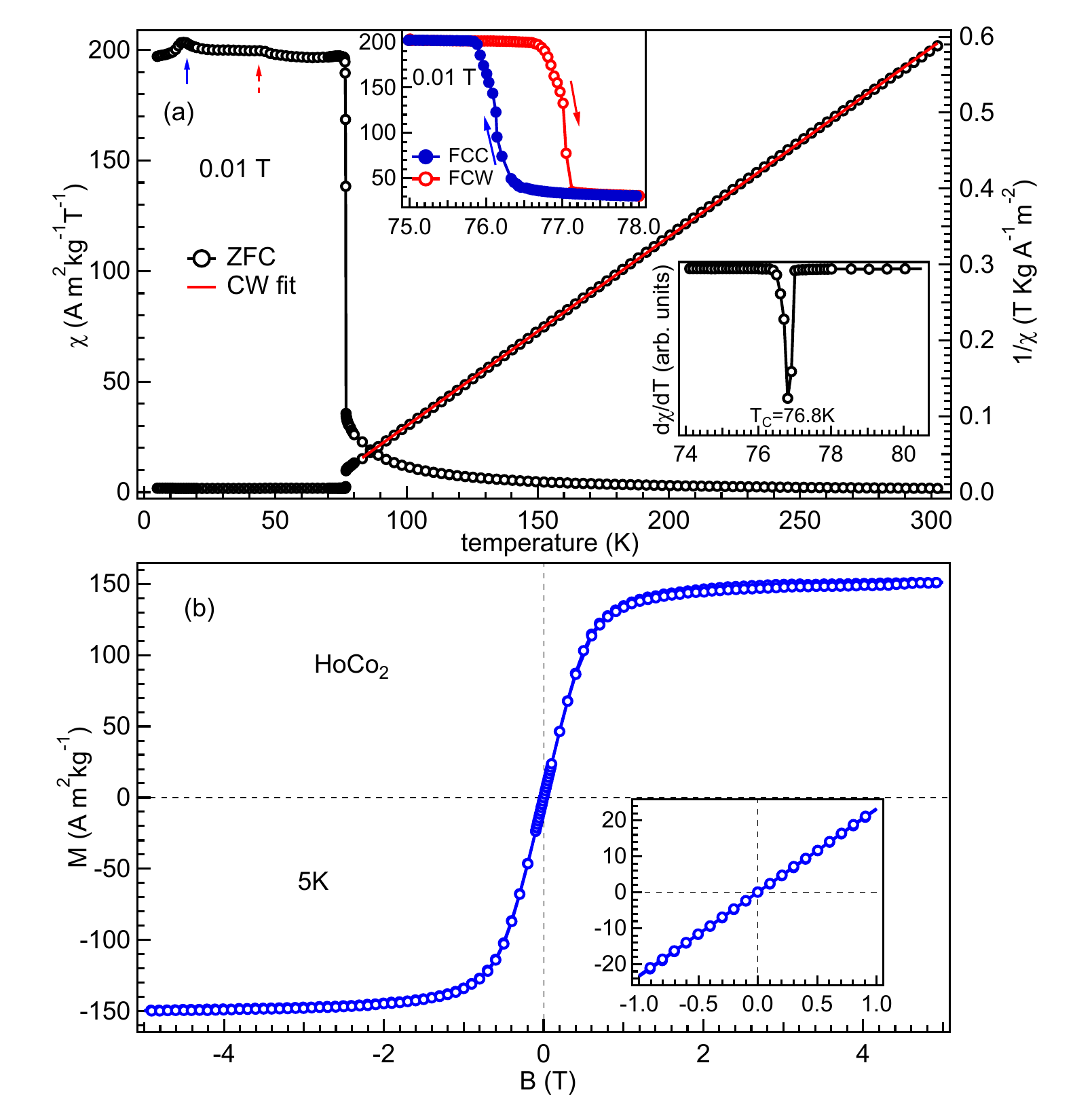} 
\caption {(a) The temperature-dependent direct (left axis) and inverse (right axis) magnetic susceptibility ($\chi$-T) of HoCo$_2$ measured in the temperature stable zero-field-cooled (ZFC) mode at 0.01~T. The solid red line represents the best fit using the Curie-Weiss law in the PM state. The upper inset represents the $\chi$-T data recorded using the temperature sweep mode at a rate of 0.1 K/min in field-cooled cooling (FCC) and field-cooled warming (FCW) protocols in the vicinity of T$_{\rm C}$. The lower inset shows the first derivative of the susceptibility in the transition region. (b) Field-dependent magnetization at 5 K. The inset shows an enlarged view of the low-field region.}
\label{Fig1_MTMH}
\end{figure}

The temperature-dependent magnetic susceptibility ($\chi = M/H$ vs. $T$) of HoCo$_2$, measured at 0.01~T during heating (after cooling in zero field) in a temperature-stable mode, is presented in Fig. \ref{Fig1_MTMH}(a). A very sharp transition in magnetic susceptibility is observed at 76.8$\pm$0.1K [see d$\chi$/dT in the lower inset of Fig. \ref{Fig1_MTMH}(a)], which is consistent with the reported first-order ferrimagnetic (FiM) to paramagnetic (PM) transition in HoCo$_2$ \cite{Gignoux_PRB_75, Mudryk_JMCC_91, Albillos_PRB_06, Bykov_Jalcom_24}. An additional transition around 14~K (indicated by the solid blue arrow), attributed to spin reorientation, and a cusp around 45~K (indicated by the dashed red arrow), are also consistent with previous reports \cite{Mudryk_JMCC_91, Bykov_Jalcom_24, Cuong_JalCom_97, Aubert_SSC_78}. No thermal hysteresis was observed at $T_{\rm C}$ in the heating and cooling (not shown) $\chi$-$T$ curves recorded in the temperature-stable mode, consistent with Ref. \cite{Mudryk_JMCC_91}. Here, it is important to emphasize that HoCo$_2$ is known to exhibit a very sharp transition with a small thermal hysteresis ($\approx$2~K \cite{Gignoux_PRB_75}) at $T_{\rm C}$. Therefore, the $\chi$-$T$ curves recorded during heating and cooling, by stabilizing the temperature at each point, may introduce significant inaccuracies due to slight overshoots in the settled temperature, leading to additional transformations of the sample between the low- and high-temperature states. Thus, the temperature sweep mode is more appropriate in this case, where the data are recorded with a monotonic increase or decrease in temperature \cite{Guillou_NC_18}. Interestingly, the heating and cooling $\chi$-$T$ curves recorded in the temperature sweep mode clearly display a thermal hysteresis of approximately 0.9~K (recorded with a sweep rate of 0.1~K/min), as shown in the upper inset of Fig. \ref{Fig1_MTMH}(a), confirming the first-order nature of this transition. However, this value is still lower than the 2~K reported in Ref. \cite{Gignoux_PRB_75}. Note that unlike specific heat measurements, the instantaneous temperature of the sample during magnetization measurements in the SQUID magnetometer is not directly monitored or regulated. Instead, temperature control is mediated indirectly through an exchange gas, leading to a time lag before the sample reaches thermal equilibrium. Consequently, as the temperature sweep rate increases, thermal hysteresis also rises. Therefore, measurements conducted at varying sweep rates are essential to accurately extrapolate the thermal hysteresis at a near-zero sweep rate \cite{Guillou_NC_18}.  Furthermore, the temperature-dependent inverse susceptibility is shown on the right axis of Fig. \ref{Fig1_MTMH}(a), which was fitted in the 80-300~K range using the Curie-Weiss (CW) law, $\chi = \frac{C}{T - T_\theta}$, where $C$ and $T_\theta$ are the Curie-Weiss constant and Curie temperature, respectively, as indicated by the solid red line. The best-fit curve gives $T_\theta = 65$~K and an effective magnetic moment ($\mu_{\rm eff}$) of 9.53 $\mu_B$/f.u. The slightly lower value of $T_\theta$ compared to $T_{\rm C} \approx 77$~K is characteristic of very sharp first-order transitions, while the smaller value of $\mu_{\rm eff}$, compared to the theoretical value for the free Ho$^{3+}$ ion with the $^5$I$_{8}$ ground state (10.61$\mu_B$), may be attributed to crystal field effects in the lattice.\par

In Fig. \ref{Fig1_MTMH}(b), we present the field-dependent magnetization measurements at 5~K. We observe negligible coercivity in the sample, as can be seen from the enlarged view of the low field region presented in the inset of Fig. \ref{Fig1_MTMH}(b). This makes HoCo$_2$ a special case with very small thermal as well as magnetic hysteresis and a large change in the magnetization at $T_{\rm C}$, desirable for magnetocaloric applications \cite{Nehan_PCCP_24}. Furthermore, the magnetic moment increases rapidly up to around 2~T and then increases very slowly up to the highest measured field of 5~T. The magnetic moment at 5~T is 7.6~$\mu_B$/f.u., which is consistent with the reported value of 7.7~$\mu_B$/f.u. along $<$110$>$ at 4.2~K, resulting from the antiparallel alignment of the 9.3~$\mu_B$ Ho and -0.8 $\mu_B$ Co moments \cite{Gignoux_PRB_75}.\par

 \begin{figure*}
\includegraphics[width=7in]{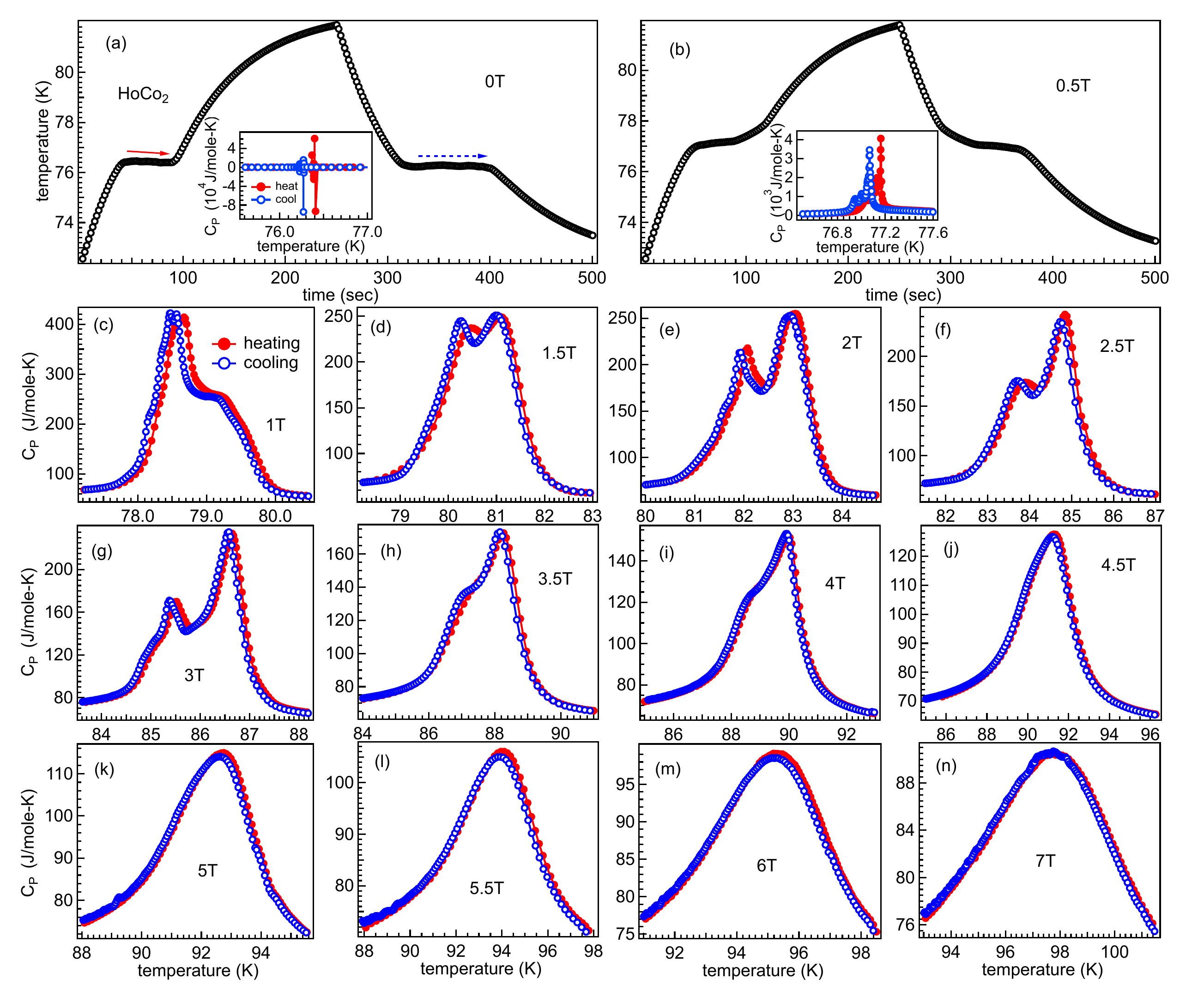} 
\caption {(a) The time dependence of the sample temperature during heating (solid red arrow) and cooling (blue dashed arrow), after the application and removal of a fixed heat pulse, respectively, at 0~T and (b) 0.5~T magnetic fields. The insets show the resulting specific heat (C$_{\rm P}$) curves of the sample during both heating and cooling cycles. (c--n) The temperature-dependent C$_{\rm P}$ data of HoCo$_2$ in the vicinity of the magnetic transition at different magnetic fields.}
\label{Fig2_HC}
\end{figure*}

 \begin{figure}
\includegraphics[width=3.5in]{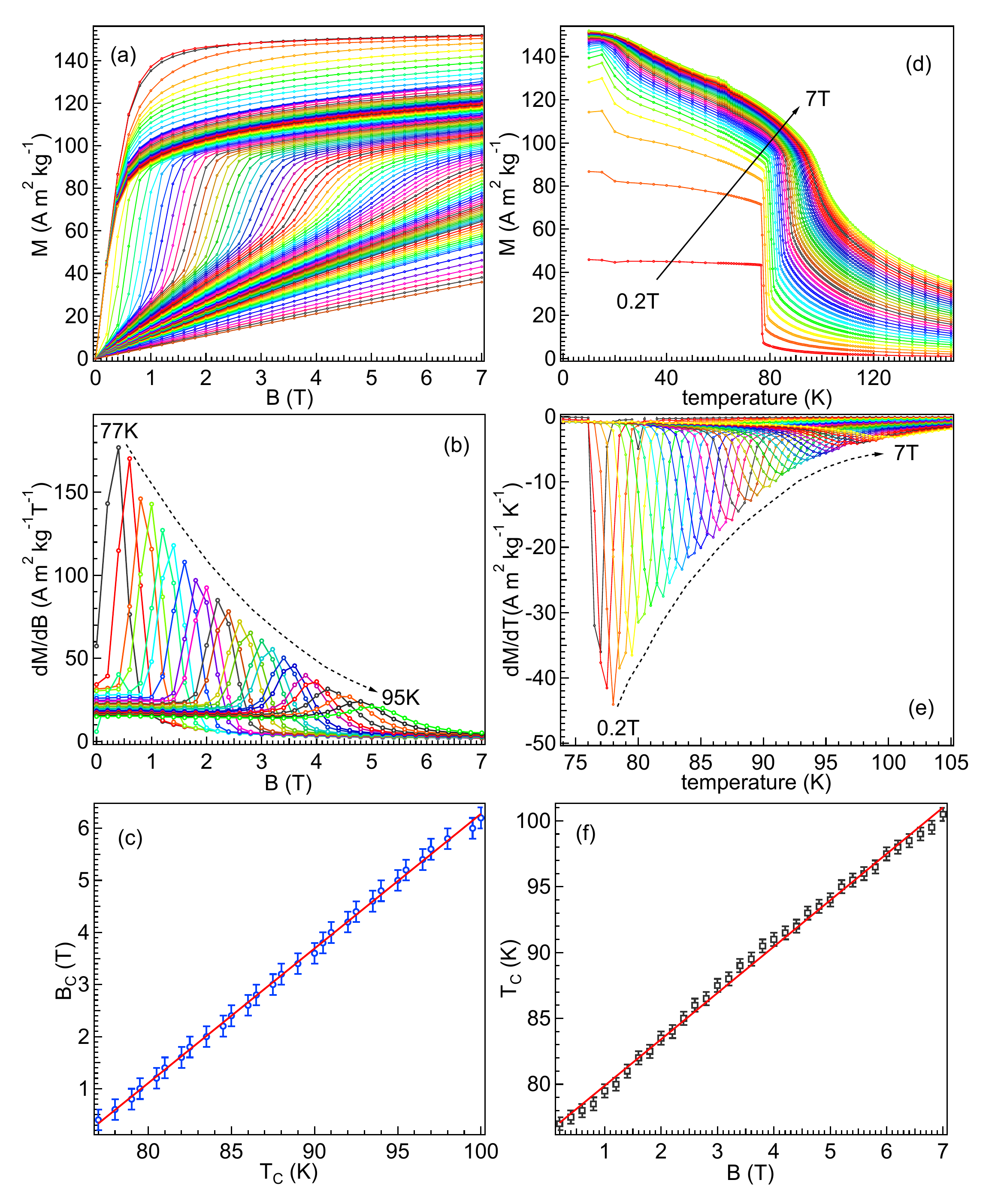} 
\caption {(a) Virgin magnetization isotherms of HoCo$_2$ from 0 to 7~T across T$_{\rm C}$. (b) The magnetic field derivative of the magnetization from 77 to 95~K. (c) The temperature dependence of the critical magnetic field. The red solid line represents the linear fit of the data. (d) The temperature-dependent magnetization at different fields, extracted from the virgin magnetization isotherms. (e) The temperature derivative of the magnetization at different fields. (f) Field dependence of the transition temperature, where the red line represents the linear fit. }
\label{Fig3_virgin}
\end{figure}

To further understand the behavior of the FiM to PM transition in HoCo$_2$ in response to an external magnetic field, we performed specific heat (C$_{\rm P}$) measurements from 0 to 7~T in the vicinity of $T_{\rm C}$. The specific heat measurements performed using the conventional 2$\tau$ relaxation method assume that the value of C$_{\rm P}$ remains constant during a given heat pulse and has the same value during the heating and cooling of the sample. However, for very sharp first-order transitions, as in the present case, these assumptions are not valid, leading to inaccurate values of C$_{\rm P}$, which obscure the true thermodynamic nature of such transitions \cite{Hardy_JPCM_09, Lashley_Cryo_43}. Therefore, a point-by-point time-dependent analysis of both heating and cooling curves separately, after applying a long heat pulse  which can capture the whole transition, is essential to precisely estimate the C$_{\rm P}$ in the vicinity of FO transitions  \cite{Lashley_Cryo_43, Kumar_PRB_24}. We performed specific heat measurements of HoCo$_2$ using the long heat pulse method to accurately probe the sharp features of the sample and their evolution with the magnetic field. \par

Figure \ref{Fig2_HC}(a) shows the time evolution of the sample temperature ($t$-$T$) in the vicinity of the phase transition during heating (indicated by the solid red arrow) and cooling (indicated by the dashed blue arrow) after applying and removing a long heat pulse to the sample at zero field. Interestingly, both heating and cooling curves exhibit a flat temperature region where the sample temperature does not increase (decrease) even when the sample heater was on (off). This clearly indicates the presence of a large latent heat in the system, confirming the first-order nature of this magnetostructural phase transition. We observe a thermal hysteresis of around 0.15~K between the heating and cooling curves, which is smaller than the hysteresis observed in the magnetization measurements discussed earlier.  This discrepancy arises because, unlike in magnetization measurements, the sample is in direct contact with both the thermocouple and heater during specific heat measurements, allowing precise control and monitoring of the sample's instantaneous temperature.  In the absence of a magnetic field, this constant temperature region in the $t$-$T$ curves results in a very large value of $C_{\rm P}$ at $T_{\rm C}$ [virtually infinite; see inset of Fig. \ref{Fig2_HC}(a)]. We also observe a large negative specific heat due to small fluctuations in the sample temperature in the vicinity of the first-order transition \cite{Piazzi_PRA_17, Erbesdobler_JAP_20, Bennati_APL_20, Guillou_SM_19, Hardy_AM_22}.  With the application of the magnetic field, the $t$-$T$ curves split into two distinct slopes, resulting in two corresponding peaks in the $C_{\rm P}$ curves, as shown in Figs. \ref{Fig2_HC}(b--n). This effect is clearly visible for $B \geqslant 1$~T [see Figs. \ref{Fig2_HC}(c--n)]. These results are in nice agreement with those recorded using the conventional 2$\tau$ relaxation technique in high resolution, as compared in Fig. 1 of supplementary material for B=2~T. The two peaks are separated by approximately 1.5~K, and the separation between them remains constant as the magnetic field increases. However, the strength of the LT peak decreases relative to the HT peak as the magnetic field increases. The broadening of these features also increases with the magnetic field, and for $B \geqslant 5$~T, the two features merge into a single broad peak [see Figs. \ref{Fig2_HC}(k--n)]. We used more than one heat pulse to capture the wide temperature range for B $\geqslant$ 5~T. \par

A two-peak feature in the differential scanning calorimetry (DSC) results for HoCo$_2$ was also observed in Ref. \cite{Albillos_PRB_06} in the low-field region ($B < 5$~T), despite the isotropic nature of the transition temperature in this compound \cite{Gignoux_PRB_75}. The authors attributed the two peaks to the presence of a small amount of off-stoichiometric component in the sample with a different transition temperature \cite{Gignoux_PRB_75}. However, XRD and EDS measurements performed on the present sample, as well as in Ref. \cite{Albillos_PRB_06}, show no traces of compositional inhomogeneity or a secondary phase in the compound within the instrumental limits. The sharp single transition in the absence of a magnetic field further discard the possibility of two different phases as the origin of the observed doublet in the C$_{\rm P}$(T) curves in the present case.  The first-order nature of the transition diminishes with increasing magnetic field: a significant decrease in the strength and an increase in the broadening of the C$_{\rm P}$ curves (see Fig. 2 of supplementary material), accompanied by the continuous change in the $t$-$T$ curves (see Fig. 3 of supplementary material) and a smaller thermal hysteresis at T$_{\rm C}$, have been observed. These observations point toward a possible field-induced transition from a first-order to a second-order nature of this phase transformation in HoCo$_2$. However, the continuity in the $t$-$T$ curves and the broadening of the C$_{\rm P}$ curves are subjective indicators. Therefore, in-depth magnetization measurements are necessary to clearly determine the thermodynamic nature of this magnetic transition in HoCo$_2$ at higher magnetic fields.\par

The high-precision virgin magnetization isotherms of HoCo$_2$ were recorded from 0 to 7~T across T$_{\rm C}$, as presented in Fig. \ref{Fig3_virgin}(a). For $T \leqslant T_{\rm C}$, the curves exhibit saturating behavior similar to that observed at 5~K; however, for $T > T_{\rm C}$, the magnetic moment increases almost linearly for lower applied magnetic fields and then shows a step-like jump with further increases in $B$ [see Fig. \ref{Fig3_virgin}(a)]. The critical magnetic field for this jump in magnetization, i.e., the magnetic field required to ferrimagnetically align the spins, increases with temperature. To clearly illustrate this, we present the magnetic field derivative of the magnetization (dM/dB) in Fig. \ref{Fig3_virgin}(b) for $T > T_{\rm C}$, where a monotonic shift in the peak position toward higher fields is clearly observed with increasing sample temperature. We perform Gaussian fitting of the peaks near the transition to precisely estimate the peak position. In Fig. \ref{Fig3_virgin}(c), we plot this peak position as a function of temperature, which shows a linear increase for $T > T_{\rm C}$.  A straight fit of the experimental data, represented by the red solid line in Fig. \ref{Fig3_virgin}(c), reveals a significant shift of 0.26~T/K. \par

The temperature-dependent magnetization extracted from the virgin magnetization isotherms is shown in Fig. \ref{Fig3_virgin}(d) for different magnetic fields. With increasing magnetic field, the transition becomes broader, and at B =7 ~T, it become almost continuous, suggesting that the magnetic field may have transformed the first-order FiM$\rightarrow$PM transition into a second-order type. The temperature derivative of the magnetization (dM/dT) is plotted as a function of temperature in Fig. \ref{Fig3_virgin}(e) for different fields to illustrate the field evolution of $T_{\rm C}$. The field dependence of $T_{\rm C}$ is shown in Fig. \ref{Fig3_virgin}(f), exhibiting linear behavior up to 7~T with a shift of 3.5~K/T, which is close to the values reported in \cite{Mudryk_JMCC_91} (3.3~K/T) and \cite{Bykov_Jalcom_24} (4.0~K/T). This linear dependence of the transition temperature of HoCo$_2$ on the external field makes it a suitable candidate for device applications across a wide range of operating temperatures. Interestingly, dM/dT exhibits a single and symmetric peak at all fields, and its peak position almost coincides with the high-temperature peak of the C$_{\rm P}$ curves (see Fig. 4 of supplementary material), suggesting that a non-magnetic aspect may be associated with the low-temperature peak of the C$_{\rm P}$ curves. However, a simultaneous study of the magnetic and structural changes is necessary to probe the origin of two peaks within a narrow temperature range \cite{Aubert_IEEE_22, Karpenkov_APR_23}.\par

 \begin{figure}
\includegraphics[width=3.5in]{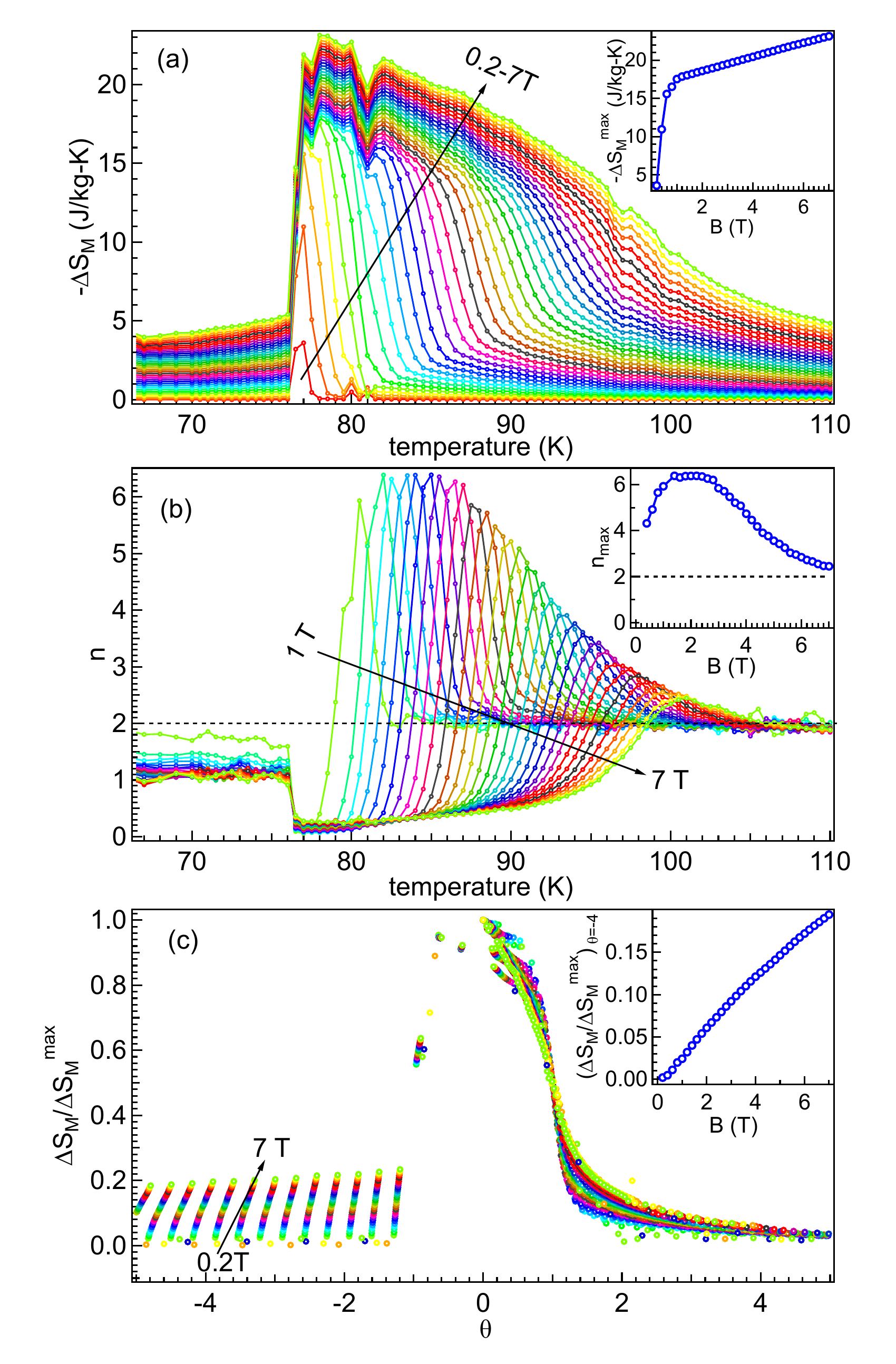} 
\caption {(a) The temperature-dependent magnetic entropy change ($\Delta$S$_M$) of HoCo$_2$ at different magnetic fields. The inset shows the field dependence of the maximum magnetic entropy change ($\Delta$S$^{\rm max}_M$). (b) The temperature-dependent local field exponent ($n$) of $\Delta$S$_M$ at different fields. The inset shows the field dependence of the maximum value of $n$ ($n_{\rm max}$) in the vicinity of T$_{\rm C}$. (c) The normalized $\Delta$S$_M$ versus scaled temperature curves at different fields. The inset shows the field dependence of the vertical dispersion at $\theta = -4$.}
\label{Fig4_delS}
\end{figure}

The specific heat and M-T measurements qualitatively suggest a change in the nature of the FiM to PM transition in HoCo$_2$ from first-order to second-order at higher magnetic fields. Therefore, to quantitatively understand this, we calculate the temperature-dependent magnetic entropy change [$\Delta S_M$(T)] for different $\Delta H$ values using the following Maxwell thermodynamic relation \cite{Phan_JMMM_07, Kumar_PRB1_20}: 

\begin{eqnarray} 
\Delta S_M(T,H)=\int_0^H \left(\frac{\partial M(T,H)}{\partial T}\right)_H dH 
\end{eqnarray} 

The $\Delta$S$_M$(T) curves for HoCo$_2$ are shown in Fig. \ref{Fig4_delS}(a), exhibiting a giant magnetocaloric effect with a maximum magnetic entropy change ($\Delta S_M^{\rm max}$) of 18.5~J/kg-K at $\Delta $H = 2~T. The $\Delta S_M(T)$ curves remain almost symmetric up to $H=1 $~T and then gradually become asymmetric towards the PM region with further increases in the magnetic field [see Fig. \ref{Fig4_delS}(a)]. It is interesting to note that the the value of $\Delta S_M^{\rm max}$ increases rapidly up to $\sim 1$~T and then increases at a much slower rate, as shown in the inset of Fig. \ref{Fig4_delS}(a). Recently, J. Y. Law \textit{et al.} proposed a general criterion to distinguish between first- and second-order phase transitions based solely on magnetization measurements \cite{Law_NC_18}. The authors proposed that the power exponent ($n$) of the field dependence of magnetic entropy change ($\Delta S_M \propto H^n$) exhibits a value greater than 2 only for first-order phase transitions \cite{Law_NC_18}. Therefore, we calculate the local field exponent of entropy, $n$, defined as: 

\begin{eqnarray} 
n(H, T)= \frac{d(\ln|\Delta S_M|)}{d(\ln H)} 
\end{eqnarray} 

The $n$(T) values for HoCo$_2$ are shown in Fig. \ref{Fig4_delS}(b) for $\Delta$H =1-7~T. Note that for second-order FM to PM transitions, $n=1$ in the FM state, attaining a minimum value at $T_{\rm C}$, given by $n(T_{\rm C})=1 + (1 - \beta^{-1}) \delta^{-1}$, where $\beta$ and $\delta$ are the critical exponents, followed by $n=2$ in the PM state, in accordance with the Curie-Weiss law \cite{Franco_APL_06}. In contrast, $n > 2$ represents the first-order nature of the magnetic transition \cite{Law_NC_18}. Interestingly, in the present case, $n$ increases monotonically in the vicinity of $T_{\rm C}$, reaches a maximum value of $n \sim 6.5$ for $H \approx 1.2$~T, and then decreases with further increases in the magnetic field [see Fig. \ref{Fig4_delS}(b)]. Notably, in the vicinity of $T_{\rm C}$, $n > 2$ even at $\Delta $H=7~T, indicating that the first-order nature of this phase transition in HoCo$_2$ persists even up to the highest applied magnetic field (7~T). The inset of Fig. \ref{Fig4_delS}(b) shows the field dependence of the maximum value of the exponent $n$ ($n_{\max}$), which clearly indicates that $n > 2$ for $\Delta $H = 7~T [see Fig. 5(a) of the supplementary material for more clarity]. \par

Additionally, a criterion based on the extent of the collapse of all the $\Delta S_M$(T) curves into a universal master curve (UMC), after properly rescaling the $\Delta S_M$ and temperature axes, has been proposed to distinguish the order of the phase transition \cite{Bonilla_PRB_10}. For a second-order phase transition, curves at different fields collapse into a single UMC across the entire temperature range, whereas a vertical dispersion is observed for first-order transition materials below $T_{\rm C}$ \cite{Bonilla_PRB_10}. Therefore, we construct a UMC of $\Delta S_M(T)$ at different fields by plotting scaled entropy ($\Delta S_M / \Delta S_M^{\rm max}$) versus scaled temperature ($\theta$), defined as: 

\begin{eqnarray} 
\theta= 
\begin{cases} 
 -(T - T_{pk}) / (T_{r1} - T_{pk}) & T \leqslant T_{\rm C} \\
 (T - T_{pk}) / (T_{r2} - T_{pk}) & T > T_{\rm C}
\end{cases}
\end{eqnarray} 

where $T_{r1}$ and $T_{r2}$ are two reference temperatures below and above $T_{\rm C}$, respectively, for a certain value of $h = \Delta S_M / \Delta S_M^{\rm max}$, and $T_{pk}$ represents the temperature corresponding to $\Delta S_M^{\rm max}$ \cite{Franco_APL_06}. The normalized entropy versus scaled temperature for HoCo$_2$ is presented in Fig. \ref{Fig4_delS}(c) for different applied magnetic fields for $h = 1/2$. We observe significant dispersion in the scaled entropy curves even up to 7~T, primarily below $T_{\rm C}$. To clearly present this, we show the field dependence of the normalized entropy at $\theta = -4$ in the inset of Fig. \ref{Fig4_delS}(c), where a monotonic enhancement in the dispersion of the scaled curves, even up to 7~T, confirms the stability of the first-order nature of this phase transition in HoCo$_2$ in response to the external magnetic field [see Fig. 5(b) of the supplementary material for more clarity].

 \begin{figure}
\includegraphics[width=3.5in]{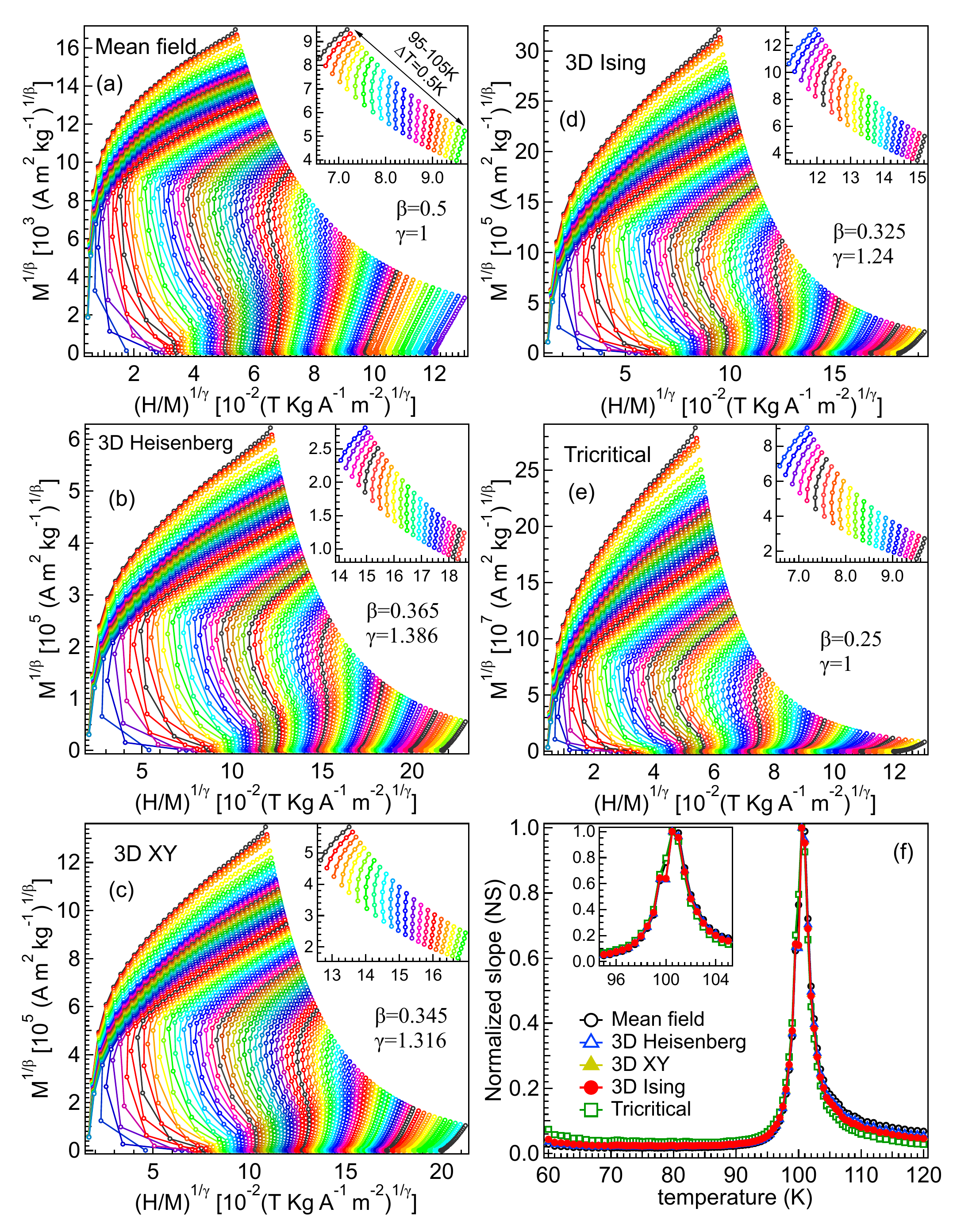} 
\caption {(a--e) The modified Arrott plots (MAPs) for different universality classes across the magnetic transition temperature. The inset shows a magnified view for the 6--7~T magnetic field and the 95--105~K temperature range. (f) The temperature-dependent normalized slope (NS) of the MAPs for different universality classes. The inset shows an enlarged view from 95 to 105~K. }
\label{Fig5_Arrott}
\end{figure}

 Here, it is essential to highlight that the $\Delta S_M(T)$ curves calculated at higher magnetic fields may retain contributions from the low-field data (refer to Eq. 1). Consequently, even when the magnetic transition shifts from first-order to second-order at elevated fields, the influence of the low-field data, where the transition is unequivocally first-order in nature, can affects the values of $\Delta S_M(T)$ at higher fields, compared to a purely second-order transition. This may contribute to an overshoot in the power exponent $n$ and dispersion of the $\Delta S_M$ curves in the high-field region. Therefore, to further investigate this, we apply the Banerjee Criterion, which states that a negative (positive) slope in the Arrott plots (H/M vs. M$^2$) signifies a first- (second-) order phase transition \cite{Banerjee_PL_64, Arrott_PRL_67}. In Fig. \ref{Fig5_Arrott} (a), we present the Arrott plots over a wide temperature range from 60 to 120~K. A negative slope in the Arrott curves can be clearly observed even for H $\geqslant$ 6~T. In fact, the slope of the Arrott curves changes from positive to negative across T$_{\rm C}$ in the 6-7~T range, as shown more clearly in the inset of Fig. \ref{Fig5_Arrott} (a), which indicates the first-order nature of the transition even at high magnetic fields. However, it is important to note that the Banerjee Criterion is only valid for materials following the mean-field theory with critical exponents $\beta$=0.5 and $\gamma$=1. Consequently, we construct modified Arrott plots (MAPs) [M$^{1/\beta}$ vs. (H/M)$^{1/\gamma}$] using virgin magnetization isotherms for different universality classes with their respective critical exponents \cite{Guillou_PRB_80}, as shown in Figs. \ref{Fig5_Arrott}(b--e). For the correct choice of critical exponents in the case of a second-order phase transition, the Arrott curves are expected to display linear behavior parallel to each other in the high-field region \cite{Sarkar_PRB_08, Biswas_PRM_24, Roble_PRB_04}. A similar approach has also been utilized to understand the field-induced first-order to second-order phase transition in Sm$_{0.52}$Sr$_{0.48}$MnO$_3$, where MAPs show straight curves for H$ > $4~T, indicating a transition to second-order \cite{Sarkar_PRB_08}. Interestingly, in our case, none of the universality classes exhibit linear behavior in the Arrott curves near the transition temperature, even for H$ > $6~T [see Figs. \ref{Fig5_Arrott}(b--e) and their insets]. The clear curvature in these curves is a characteristic feature of the first-order nature of this phase transition up to 7~T. We attempted to fit these curves linearly in the 6-7~T range (see Fig. 6 of the supplementary material) and plotted the normalized slopes (slope at a given temperature/slope at T$_{\rm C}$) of these curves in Fig. \ref{Fig5_Arrott}(f). A sharp peak in the slope at T$_{\rm C}$ is clearly observed, even within a narrow temperature range [see the inset of Fig. \ref{Fig5_Arrott}(f)]. This indicates that the critical exponents of the sample diverge at T$_{\rm C}$, further suggesting the persistence of the first-order nature of this transition at least up to 7~T.

\section{\noindent ~Conclusion}

A detailed investigation of specific heat and magnetization as functions of both temperature and applied magnetic fields has been conducted on polycrystalline HoCo$_2$. We observe significant broadening of the $C_{\rm P}$ peak, accompanied by a reduction in its magnitude and thermal hysteresis at $T_{\rm C}$ as the magnetic field increases. This, along with the gradual change in magnetization at $T_{\rm C}$ at elevated fields, points towards a transformation from a first- to a second-order nature of the FiM to PM transition in HoCo$_2$.   However, the behavior of the field exponent of magnetic entropy, $n$, and the dispersion in the scaling behavior of $\Delta S_{\rm M}$ curves suggest the persistence of the first-order character of the phase transition up to at least 7~T. The first-order nature of the transition is also indicated by the non-linearity in the MAPs, even for fields of 6--7~T. These findings on the nature of the magnetic transition in HoCo$_2$ under strong magnetic fields contribute valuable insights for its application in magnetocaloric technologies. Further, a sharp, single peak in the $C_{\rm P}$ curves at zero field splits into two components in the presence of a magnetic field, with their relative strengths varying with the field. Moreover, using HoCo$_2$ as an example, we demonstrate the importance of performing specific heat and magnetization measurements in sweep mode, i.e., continuously increasing or decreasing the sample temperature across the phase transition, to accurately probe the material's response in the vicinity of first-order transitions, particularly those with small thermal hysteresis.

\section*{\noindent ~Supplementary material}

See the supplementary material for the details of the specific heat and magnetization data. 

\section*{Acknowledgments}

This work was performed at Ames National Laboratory and was supported by the Division of Materials Science and Engineering of the Office of Basic Energy Sciences, Office of Science of the U.S. Department of Energy (DOE). Ames National Laboratory is operated for the U.S. DOE by Iowa State University of Science and Technology under Contract No. DE-AC02-07CH11358. 

\section*{Data Availability}
The data that support the findings of this study are available from the corresponding author upon reasonable request.

\end{document}